\documentclass[runningheads]{llncs}
\pdfoutput=1

\usepackage{graphicx}
\usepackage{cite}

\usepackage[linesnumbered,ruled,lined]{algorithm2e}
\usepackage{algpseudocode}
\usepackage{booktabs}

\usepackage{tabularx}
\usepackage{array}
\usepackage{multirow, makecell}
\usepackage{multicol}
\usepackage{arydshln}
\usepackage{pifont}
\usepackage{xcolor}
\usepackage{color,soul}
\usepackage[misc,geometry]{ifsym}
\begin{document}
\title{Security and Privacy Issues and Solutions in Federated Learning for Digital Healthcare}

%

\author{Hyejun Jeong {[\Letter]} \and Tai-Myoung Chung* {[\Letter]}}

\authorrunning{H.Jeong}
\titlerunning{Security and Privacy Issues and Solutions in FL for Digital HC}

%
\institute{College of Computing, Sungkyunkwan University, \\ Suwon, Korea (Republic of)\\
\email{\{june.jeong\}@g.skku.edu} and \email{\{tmchung\}@skku.edu}}
\maketitle              

\begin{abstract}
The advent of Federated Learning has enabled the creation of a high-performing model as if it had been trained on a considerable amount of data. A multitude of participants and a server cooperatively train a model without the need for data disclosure or collection. The healthcare industry, where security and privacy are paramount, can substantially benefit from this new learning paradigm, as data collection is no longer feasible due to stringent data policies. Nonetheless, unaddressed challenges and insufficient attack mitigation are hampering its adoption. Attack surfaces differ from traditional centralized learning in that the server and clients communicate between each round of training. In this paper, we thus present vulnerabilities, attacks, and defenses based on the widened attack surfaces, as well as suggest promising new research directions toward a more robust FL.

\end{abstract}

\section{Introduction}
\label{section1}

Digital health has rapidly grown, and the COVID-19 outbreak accelerated its evolution. However, HIPAA reported that there were 712 healthcare data breaches in 2021, exceeding 2020 by 11\% \cite{hipaa_journal_2022}, and Verizon confirmed that data breaches in the healthcare industry increased by 58\% during the pandemic. An Electronic Health Record (EHR) contains a wealth of sensitive private information about each patient, such as name, social security number, financial information, current and previous addresses, and medical history. However, traditional digital healthcare relies on centralized AI techniques that operate on a single location such as a server or data center for analytics; thus, it requires data collection. It is often not only time and resource-consuming but likely to violate stringent privacy protection policies, such as GDPR, CCPA, and HIPAA, that mandate securing patient health-related data management. 

Federated Learning (FL) decouples the use of AI techniques from gathering data by training a global model in a distributed manner under the orchestration of a central server (often referred to as an aggregator) and multiple local clients. The server updates the global model by aggregating the local models’ parameters, trained using each end-device data. In other words, data does not leave the data-owning devices, so it reduces the risk of raw training data being exposed in the middle of communication. FL thus came to light for its data privacy improvement, allowing learning without data leakage in situations where personal information must be protected. The advantages hold great promise to leverage AI techniques in the healthcare sector while complying with privacy policies. 

Despite the benefits, FL opens new attack surfaces and vulnerabilities that adversaries can exploit to harm the global model or leak the data. The expanded attack surfaces necessitate an updated vulnerability analysis to minimize the threat probabilities considering the rapidly growing needs of digital healthcare. This paper aims to introduce various vulnerabilities, attacks, and defenses, as well as open challenges and future directions toward more robust federated learning in the healthcare industry.

The rest of the paper is structured as follows: Section II identifies vulnerabilities, Section III categorizes attack methods by exploiting them, and Section IV introduces defense methods. Section V suggests future research directions, and finally, we conclude the paper in Section VI.

\section{Sources of Vulnerabilities}
\label{section3}

\begin{figure}[h]
\vspace{-1.2em}
\centering
\includegraphics[width=\textwidth]{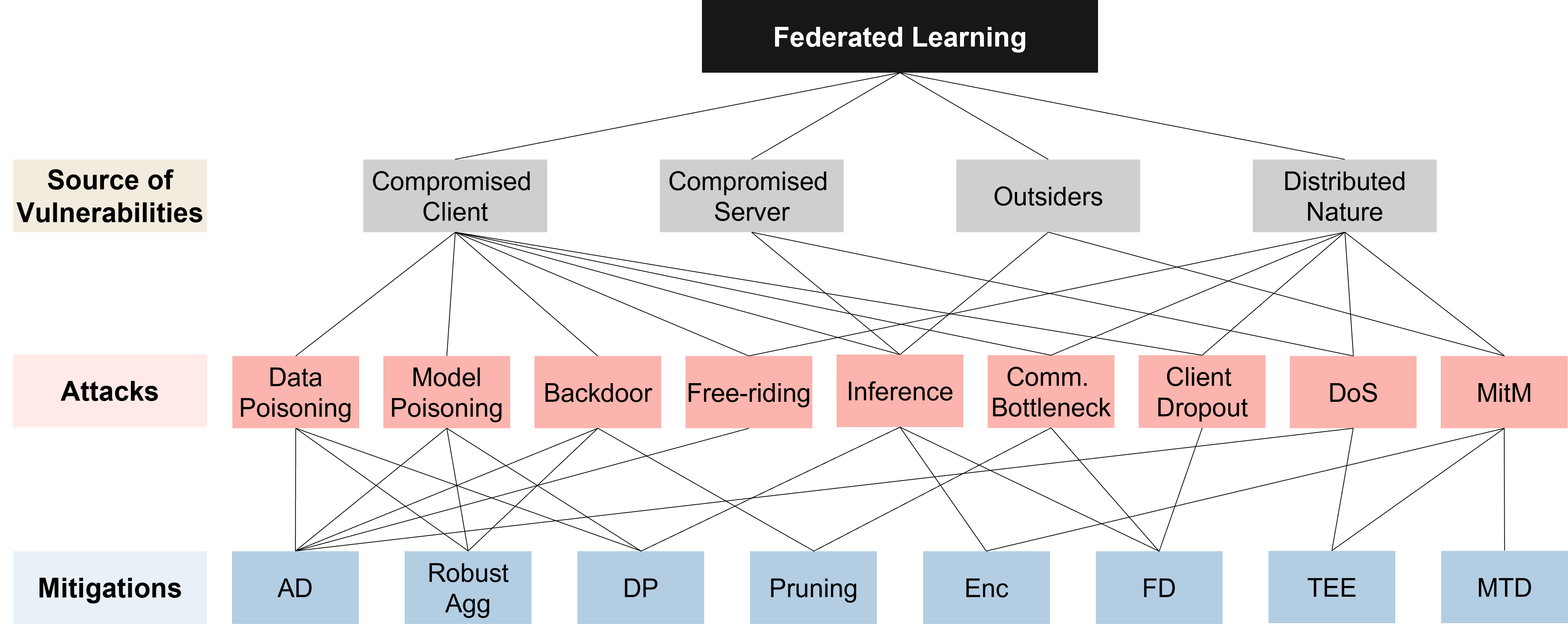}
\caption{A Taxonomy of Federated Learning: Vulnerabilities, Attacks, and Defenses.}
\label{fig:taxonomy}
\vspace{-1em}
\end{figure}

Identifying vulnerabilities in a system helps to mitigate and prevent potential attacks. FL creates a shared model by aggregating locally computed updates using client-specific data. FL involves three entities in \textbf{a distributed nature}: \textbf{clients} in which each of them computes a local model based on its own dataset, \textbf{a server} that aggregates the local model updates to recompute the latest global model, and \textbf{communication channels} that the clients and the server communicate. Fig. \ref{fig:taxonomy} visualizes the relationship between the vulnerabilities, attacks, and defense strategies in a vanilla FL framework. \\[-5pt]

\noindent\textbf{Compromisable Clients}
The FL workflow involves multiple clients computing local models based on each site-specific dataset inaccessible to other clients or the server. Thus, each client is a source of vulnerability. \textit{Compromised clients} entail the following: \vspace{-.7em}
\begin{itemize}
    \item clients unable to keep a stable connection to the server
    \item clients equipped with insufficient computing resources for a model training
    \item clients having insufficient quality or quantity of data for a fair contribution
    \item clients with malicious intents to disturb the process or to plant backdoors
\end{itemize} \vspace{-.5em}

\noindent\textbf{A Compromisable Server} or an honest-but-curious server may attempt to infer the training data from the updates, alter the model parameters, and manipulate the aggregation algorithm \cite{bouacida2021vulnerabilities}. The server is also vulnerable to flooding endangering its availability. \\[-5pt]

\noindent\textbf{Outsiders or Eavesdroppers} intercept communications between the participating parties and steal the model parameters. They also can launch consecutive attacks, such as inference or MitM attacks, using the stolen information. The outsiders also might theft the final model parameters at the deployment phase to launch inference-time attacks. \\[-5pt]

\noindent\textbf{A Distributed Nature} opens attack surfaces related to communication, such as free-riding, DoS, MitM attacks, or communication bottlenecks. The participating clients' not uniform data distribution and resources have additionally introduced non-malicious failure, resulting in detrimentally impacted model performance and extended training time.
\section{Attacks and Threats}
\label{section4}

\begin{table}[ht]
\centering
\renewcommand*{\arraystretch}{1.05}
\begin{tabular}{p{0.21\linewidth} | p{0.12\linewidth} | p{0.64\linewidth}}
\textbf{Threats}         & \textbf{Severity}          & \textbf{Description}  \\ \hline\cline{1-3}
Poisoning (\textsection\ref{subsec:poisoning})     &  High to Medium   & alter the training data or model parameters to modify the model's behavior in a malicious direction   \\ \hline
Inference (\textsection\ref{subsec:inference})     &  High to Medium   & analyze the global or local model parameters to infer the information in the training dataset   \\ \hline
Backdoor (\textsection\ref{subsec:backdoor})       &  High             & insert hidden backdoor to train the global model on malicious tasks while the main tasks are not affected    \\ \hline
Comm Bottleneck (\textsection\ref{subsec:comm})    &  High             & congested communication due to the large size of the payload of the trained model parameters    \\ \hline
Free-riding (\textsection\ref{subsec:freeriding})  &  Medium           & fake contribution to effortlessly gain the global model \\ \hline
MitM (\textsection\ref{subsec:mitm})               &  Medium to Low    & steal the model parameters in-between the endpoints to breach data or launch consequent attacks    \\ \hline
DoS (\textsection\ref{subsec:dos})                 &  Low              & flood the server to harm its availability    \\ \hline
Client Dropout (\textsection\ref{subsec:dropout}) &  Low              & forceful or accidental dropping out of participating clients due to resource instability  \\  
\end{tabular}
\\[5pt]
\caption{The Severity and Short Description of Threats}
\label{tab:severity}
\vspace{-1.5em}
\end{table}

By exploiting the vulnerabilities mentioned in Section \ref{section3}, attackers aim to achieve two objectives: performance degradation and data leakage at training or testing time. Various actors attempt to harm a global model performance by disrupting its convergence, modifying the training data or the model updates, contribution-less participation, and incurring latency. The severity and short descriptions of each threat are summarized in Table \ref{tab:severity}. 

\subsection{Poisoning Attacks}  \label{subsec:poisoning}

\begin{figure}[h]
\vspace{-1em}
\centering
\includegraphics[width=\textwidth]{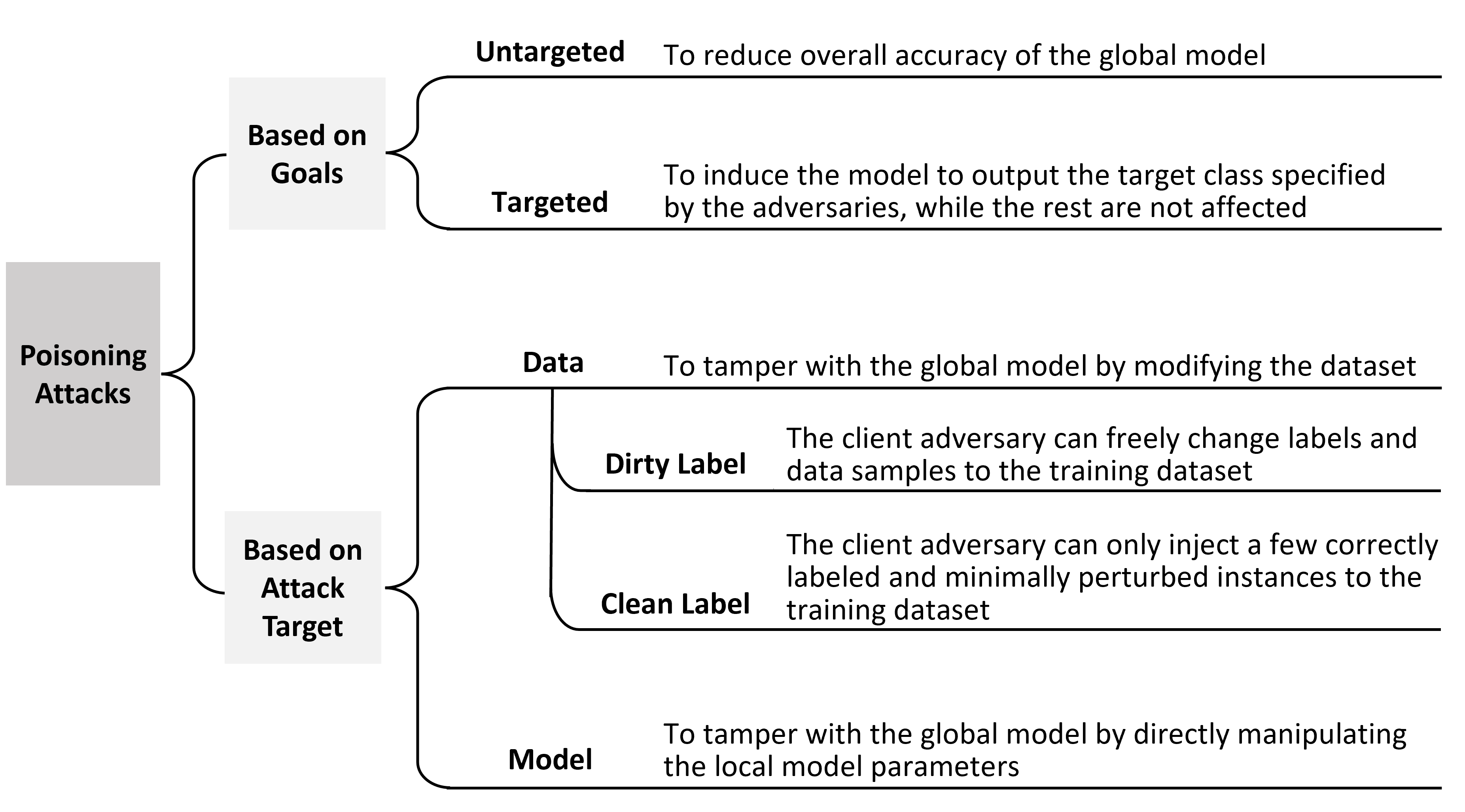}
\caption{A Taxonomy of Poisoning Attacks.}
\label{fig:poisoning}
\vspace{-.5em}
\end{figure}

FL framework is known as especially vulnerable to poisoning attacks due to its distributed and data-isolated nature. In poisoning attacks, the adversaries impact either the training dataset or local models in order to modify the behavior of the target model in some undesirable ways \cite{bouacida2021vulnerabilities}. Poisoning attacks can be categorized based on the adversarial goal: untargeted and targeted attacks. Untargeted attacks, also known as random attacks, aim to reduce the overall accuracy of a global model. On the other hand, targeted attacks aim to induce the model to output the target class specified by the adversaries, while the other classes are predicted ordinarily. Targeted attacks are generally more complicated than random ones because it has a specific goal to achieve \cite{lyu2020threats} while minimizing their influences on non-targeted classes. Fig. \ref{fig:poisoning} summarizes the categories of poisoning attacks. Based on what the adversaries attempt to manipulate, poisoning attacks are categorized into \textit{data poisoning} and \textit{model poisoning}. \\

\noindent\textbf{\textit{Data Poisoning}} attacks allow adversaries on the client side to alter the training dataset to compromise data integrity and modify the model’s behavior in an attacker-chosen direction. These attacks begin in the training phase, during local data collection \cite{mothukuri2021survey}. Data poisoning attacks fall into clean-label and dirty-label poisoning attacks. For the clean-label attacks, the adversary injects a few correctly labeled and minimally perturbed instances into the training data as if the adversarial or evasion attacks. In the FL environment, however, because the raw data instances are only observable to the data owner, dirty-label attacks prevail in which adversaries can freely change labels and data samples as they wish to misclassify \cite{bhagoji2019analyzing}. A representative dirty-label poisoning attack is a label-flipping attack, flipping the labels of two different classes to induce the global model to misclassify one class to another. \vspace{.7em}

\noindent\textbf{\textit{Model Poisoning}} attacks allow client-side adversaries to directly modify the local model parameters at the training phase before sending them to the server. The adversarial clients also manipulate the model hyperparameter, such as the learning rate, number of the local epoch, the batch size, and the optimization objective, to manipulate the training rules before the local training. The attacks improved stealthiness by optimizing the local model for both training loss and an adversarial objective to avoid deviation from the global model \cite{bouacida2021vulnerabilities, fang2020local}. Accordingly, robust aggregation rules, such as Krum\cite{blanchard2017machine}, Bulyan \cite{guerraoui2018hidden}, trimmed-mean, coomed \cite{yin2018byzantine} were proposed; however, recent works \cite{fang2020local, shejwalkar2021manipulating, wang2022flare} have broken them by tailoring model updates in a malicious direction. It has also been shown that only one non-colluding malicious client achieves targeted misclassification with 100\% confidence while ensuring the convergence of the global model \cite{bhagoji2019analyzing}. Model poisoning attacks are far more effective than data poisoning attacks because the malicious client’s updates are tuned to maximize the damage to the overall model performance while remaining stealthy \cite{bouacida2021vulnerabilities, bhagoji2019analyzing}. 

\subsection{Inference Attacks} \label{subsec:inference}
Inference attacks target participant privacy during both the training and testing phases. A compromised client, an honest-but-curious server, or outsiders may want to infer participating clients’ training data. This type of attack is not explicit to FL but has been rampant in ML or DL. A DL model contains data properties that appear unrelated to the main tasks, mainly because the gradients of a given layer are computed using this layer’s features and errors backpropagated from the layer above \cite{lyu2020threats}. Hence, trained model gradients contain extra information about the unintentional features of participants’ training data \cite{melis2019exploiting, lyu2020threats}. The attackers thus infer a substantial amount of private information such as \textit{membership}, \textit{properties}, and \textit{class representatives}. \\

\noindent\textbf{\textit{Membership Inference}} attacks are arguably the most basic privacy attack that infers the presence of a particular sample in the sensitive training dataset \cite{melis2019exploiting}. The attackers, for example, may learn whether a particular patient profile was used to train the model linked with a disease, exposing that the particular patient went to the hospital for the associated disease \cite{lyu2020threats}. Shokri et al. \cite{shokri2017membership} constructed shadow models that imitate the target model’s behavior to distinguish the target model’s output on membership versus non-members of its training dataset. Nasr et al. \cite{nasr2019comprehensive} exploit the vulnerability of the SGD algorithm, gradually influencing some parameters to adapt themselves towards reducing the loss. \\[-5pt]

\noindent\textbf{\textit{Property Inference}} attacks allow an adversarial party to infer whether the training dataset has specific general properties seemingly unrelated to the model’s primary task. For instance, a model is originally trained on facial images to predict if someone’s mouth is open (primary task); the attacker’s goal is to infer whether the training dataset is gender-balanced (inferred property) \cite{parisot2021property}. Adversarial clients can identify when a property appears and disappears in the data during the training phase by inferring from the history of global models \cite{melis2019exploiting}. \\[-5pt]

\noindent\textbf{\textit{Class Representatives Inference}} attacks allow adversarial parties to reconstruct sensitive training data features by taking advantage of their correlation with the model output \cite{zhang2020secret}. It had been demonstrated that attackers could correctly characterize features of the class from simple models, such as logistic regression and decision trees, with no false positives \cite{fredrikson2015model}. Recent efforts utilize Generative Adversarial Networks (GANs) to produce synthetic class representations from training data \cite{zhang2020secret, khosravy2022model, ho2021dp} or even recover the exact training images or texts from the gradients \cite{melis2019exploiting, geiping2020inverting, pan2020privacy}. Nevertheless, it is less feasible in FL scenarios since GAN-generated representatives are only similar to the training data when the training datasets of participants are similar (i.e., IID) \cite{melis2019exploiting}. 

\subsection{Backdoor Attacks} \label{subsec:backdoor}
Backdoor attacks include adversarial clients inserting triggers into the training data or model updates in order to train a global model on both backdoor and main tasks. On test data with the same trigger embedded, the model produces false-positive predictions with high confidence. Adversaries have also reinforced the detection evasion strategy by tuning the backdoored models to not diverge from other models \cite{wang2020attack}. For example, Bagdasaryan et al. \cite{bagdasaryan2020backdoor} rewards the model for backdoor task accuracy and penalizes it for deviating from what the aggregator considers benign. Xie et al. \cite{xie2019dba} proposed a Distributed Backdoor Attack that decomposes a global trigger pattern into separate small local patterns and embeds them into the training sets of multiple adversarial parties. Their efforts reached a higher attack success rate than inserting one global pattern. Because the malevolent behavior only appears when the triggers are present at test time, it is difficult and time-consuming to identify the existence of backdoor attacks.

\subsection{Communication Bottleneck} \label{subsec:comm}
Although FL has reduced communication costs by transmitting training models rather than much larger quantities of data, the communication overhead is still of the utmost importance, especially with larger deep learning models \cite{mothukuri2021survey}. Participating clients are typically large in number and have slow or unstable internet connections \cite{konevcny2016federated}. System heterogeneity---an inequality of computation and communication capabilities across the clients---causes asymmetric arrival timing at the server. Considering that 1 to 1000 clients participate in typical federated learning \cite{huang2020fairness}, accumulated delayed uplink time will ultimately result in a substantial delay in training time. 

\subsection{Free-Riding Attacks} \label{subsec:freeriding}
Free-riding attacks refer to an effortless extraction of a trained model; attackers benefit from the global model while not contributing to the training process. Exploiting the distributed nature and the server’s blindness, they obtain the trained global model without affording their computing resources and data \cite{lin2019free}. Free-riders generate fake updates without training with their local datasets because they have to send something to the server on each round, even if they do not update the local model parameter during the iterative federated optimization \cite{fraboni2021free}. As such, the attackers steal intellectual property and breach privacy. 

\subsection{Man-in-the-Middle attacks} \label{subsec:mitm}
In a Man-in-the-Middle (MitM) attack, the attackers position themselves in a conversation between the endpoints either to eavesdrop or to impersonate one of the participants, making it appear as if a normal exchange is taking place \cite{mallik2019man}. They aim to jeopardize confidentiality by eavesdropping, integrity by manipulation, and availability by interrupting the communication \cite{bhushan2017man}. The attackers intercept the client-server connection and replace the model parameters with malicious updates or make a shadow model to enable consecutive attacks such as model poisoning, backdoor, or inference attacks.  

\subsection{DoS attacks} \label{subsec:dos}
Denial-of-Service (DoS) attacks include an insider (compromised client) or outsider flooding a server with traffic to compromise resource availability. DoS attacks thus disrupt the server with receiving, computing, and sending model parameters. It has been a known threat in the computer network domain for a long time. For instance, Fung et al. \cite{fung2020limitations} simulated three types of DoS attacks by increasing training time, bandwidth, and CPU usage at the server and clients. As the model could not be adequately trained, making a final model will be much more expensive in computation and communication. 

\subsection{Client Dropout} \label{subsec:dropout}
Clients could be accidentally or forcefully dropped out of the procedure, which may yield ineffective results and raise concerns about fairness. Mobile devices on the client-side might frequently be offline or on slow and costly connections at any time owing to user behaviors or the unpredictable network environment in which they are located \cite{huang2020fairness}. However, because a server cannot tell whether the clients are malicious or heavily non-IID, it may exclude the clients from future rounds, thus producing a less generalized global model \cite{benmalek2022security}.

\section{Mitigation Techniques}
\label{section5}

\begin{figure}[h]
\centering
\includegraphics[width=\textwidth]{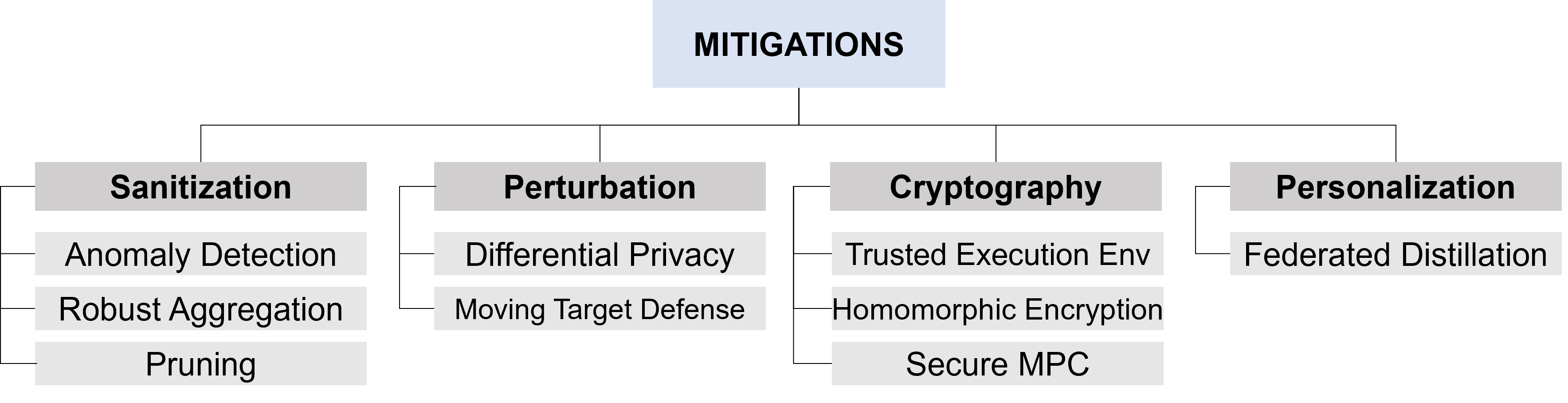}
\caption{A Category of Mitigation Techniques}
\label{fig:mitigation}
\end{figure}

A server has to secure the system by only examining the local model parameters. As illustrated in Fig.  \ref{fig:taxonomy}, a single technique can mitigate multiple issues or attacks simultaneously. A Fig. \ref{fig:mitigation} categorize the mitigation techniques. In this section, we introduce various mitigation strategies.

\subsection{Anomaly Detection} 
Anomaly or outlier detection is a proactive strategy that employs analytical and statistical methods to filter out malicious occurrences that do not conform to an expected pattern or activity \cite{mothukuri2021survey}. This technique primarily addresses poisoning, backdoor, free-riding, and DoS attacks \cite{fung2020limitations}. Fang et al. \cite{fang2020local} proposed LFR (Loss Function-based Rejection) and ERR (Error Rate-based Rejection) that reject client participation which negatively impacts the global model. Baruch et al. \cite{baruch2019little} and Sun et al. \cite{sun2019can} employed norm thresholding of client model update to remove models with boosted model parameters exceeding a specified threshold. 
ZeKoC \cite{chen2020zero} and FLAME \cite{nguyen2021flame} leverage various clustering algorithms, DBSCAN and HDBSCAN, to detect deviating clients. FoolsGold \cite{fung2020limitations} prevents Sybil-based attacks, inspired by the fact that Sybil clients’ gradients have unexpectedly high cosine similarities because they are trained for the same malicious objectives.

However, anomaly detection techniques could easily fail when clients with highly non-IID datasets join. The detection algorithm may misclassify the benign clients as abnormal because of the unique distribution of model parameters owing to learning from the non-IID dataset.

\subsection{Robust Aggregation} 
Robust aggregation is a widely studied proactive strategy to reduce the impact of malicious model updates, used as a defense technique against poisoning and backdoor attacks. According to Shejwalkar et al. \cite{shejwalkar2021manipulating}, robust aggregation algorithms (AGRs) remove the suspicious client based on the following criteria:
\begin{enumerate}
    \item Distances from the benign gradients \cite{bhagoji2019analyzing, blanchard2017machine, guerraoui2018hidden, yin2018byzantine, cao2020fltrust}
    \item Distributional differences with benign gradients \cite{bhagoji2019analyzing, sun2019can, lu2020efficient}
    \item Differences in $L_{p}$-norms of benign and malicious gradients \cite{sun2019can}
\end{enumerate}

The first criteria-based methods measure the pair-wise cosine distance between all clients \cite{cao2020fltrust, awan2021contra} or Euclidean distances between all clients and the global model \cite{blanchard2017machine, guerraoui2018hidden, yin2018byzantine} to identify malicious clients. Lu et al. \cite{lu2020efficient} followed the second criteria to address free-riding attacks; they measured the clients’ potential contributions using Gaussian Distribution. Sun et al. \cite{sun2019can} followed the second and third criteria to address label-flipping attacks. Inspired by the distinct weight distribution of malicious and benign clients, they calculated the L2 norm of local weights and compared it with a specified threshold. The server omits suspicious clients whose value is below the threshold from further aggregation. Wang et al. \cite{wang2022flare} did not directly analyze the model parameter in the parameter space. Found that Penultimate Layer Representations (PLRs) in latent space are highly differentiating features for the poisonous models, they measured the Euclidean distance of the PLRs to estimate the trust score of local models’ updates to determine the amount of weight on each local model when it comes to aggregation. 

The degree of non-IID is directly proportional to the impact of attacks because when the data distributions are highly non-IID, it is difficult for an aggregation algorithm to detect and remove the malicious clients reliably \cite{shejwalkar2021manipulating}. Regarding that IID data assumption often does not hold in practice, implementing a robust AGR without any assumptions comes as a challenge.

\subsection{Pruning} 
Pruning refers to reducing the model size by dropping neurons, thus relaxing computational complexity and communication bottleneck \cite{jiang2022model, haddadpour2021federated}, as well as addressing backdoor attacks \cite{wu2020mitigating}. Inspired by that backdoors exploit the spare capacity in the neural network, Wu et al. \cite{wu2020mitigating} dropped the backdoor attack success rate by removing spare neurons that might have been trained for backdoor tasks and fine-tuning the parameters after the training phase. 



\subsection{Differential Privacy} 
Differential Privacy (DP) introduces additional noise to the client’s sensitive data so that the attacker cannot meaningfully distinguish a single data record from the rest \cite{bouacida2021vulnerabilities}. DP is initially believed to withstand privacy attacks, such as inference attacks, but it can also inherently defend against poisoning and backdoor attacks \cite{ma2019data}. Miao et al. \cite{miao2022against} bounded the norm of malicious updates by adaptively setting a proper clipping threshold throughout the training process to eliminate backdoors and enhance the main task accuracy.

Although DP is one of the preferred techniques due to its low computational overhead and privacy quantification properties, inserted perturbation often suggests a trade-off between performance and privacy. To this extent, Nguyen et al. \cite{nguyen2021flame} estimate a minimal amount of noise to ensure the elimination of backdoors while maintaining the benign performance of the global model.

\subsection{Moving Target Defense} 
Moving Target Defense (MTD) is a proactive defense strategy against MitM attacks. Continually randomizing FL system modules obscures the vulnerability source from attackers, thereby increasing the cost and complexity of locating the exact target \cite{mothukuri2021survey}. As such, the added dynamics consequently reduce the likelihood of successful attacks and increase the system resiliency while limiting the disclosure of system vulnerabilities and opportunities for attacks \cite{mothukuri2021survey}. Zhou et al. \cite{zhou2021augmented} introduced ADS-MTD, a double-shuffle system comprising model and client-shuffling components. Specifically, MTD takes place in the second phase as a hierarchical multi-shuffler structure to dynamically and efficiently assess and eliminate malicious FL participants to enhance the aggregated model’s integrity and availability.


\subsection{Trusted Execution Environment} 
A Trusted Execution Environment (TEE) protects the FL system during a training phase by allocating a separate region for code execution and data handling. TEE guarantees integrity and confidentiality of computations while incurring lower overhead but higher privacy compared to encryption-based methods \cite{mo2019efficient}. 
TEEs, however, have limited memory for computation in order to keep the Trusting Computing Base as small as possible \cite{mo2021ppfl} to minimize the attack surface \cite{li2019teev}. Thus, Mo et al. \cite{mo2021ppfl} proposed PPFL that enlarges its memory capability by greedy layer-wise training and aggregation. They demonstrated that PPFL achieved comparable accuracy while dealing with data heterogeneity and accelerating local training processes.

\subsection{Encryption-based Methods} 


Encryption-based methods apply \textit{Homomorphic Encryption (HE)} and/or \textit{Secure Multi-party Computation (SMC)} to the model updates to combat privacy-related attacks, inference, and MitM attacks. Further, they can be combined with a perturbation-based method (i.e., DP) for more robustness.  \\[-5pt]

\noindent\textbf{\textit{Homomorphic Encryption}} enables data to be processed without decryption, and the outcome of a homomorphic operation after decryption is equivalent to the operation on plain data. Since only encrypted parameters are communicated, and the server only views and computes over the encrypted parameters, it thus can protect the data from inference attacks by a compromised server or eavesdroppers \cite{fang2021privacy}. 

However, its effectiveness comes at a substantial amount of computational, communicational, and memory overhead hampering its applicability \cite{lyu2020privacy}. Zhang et al. \cite{zhang2020batchcrypt} thus proposed BatchCrypt that sped up training time while reducing communication overhead by encoding the gradients into long integers in a batch followed by gradient-wise aggregation. \\[-5pt]

\noindent\textbf{\textit{Secure Multi-party Computation}} is a cryptographic protocol that distributes a computation process across multiple parties, with no single party having access to the data of the others. For instance, an encryption key may be divided into shares so that no individual possesses all the components needed to reassemble the key completely. SMC is a preferred approach because it is 1000 times faster than HE \cite{kanagavelu2020two}. Hao et al. \cite{hao2019efficient} and Truex et al. \cite{truex2019hybrid} integrate SMC and Gaussian-based DP to mitigate privacy threats launched by multiple colluding clients, balancing performance and privacy guarantee trade-offs by injecting a reduced amount of perturbation with the aid of SMC.

Applying SMC to FL, on the other hand, generally demands all parties to produce and exchange secret shares with all other parties. This procedure inevitably introduces a substantial communication overhead, exponentially growing with the number of participants \cite{kanagavelu2020two}.

\subsection{Federated Distillation }
Federated Distillation (FD) \cite{jeong2018communication} is a variant of the model compression techniques \cite{mothukuri2021survey} to effectively handle communication bottlenecks and heterogeneity and to combat inference attacks. FD refers to transferring knowledge from a large and fully trained teacher model to another small student model without losing validity. Sharing knowledge instead of model weights saves communication and computational costs in resource-restricted local devices as well as protects the model information from being interpreted for inference attacks \cite{wu2022communication, mothukuri2021survey}. Zhang et al. \cite{zhang2022fine} built an auxiliary generator in a server to fine-tune the model aggregation procedure. Exploiting its powerful processing capability, the server safely explores knowledge in local models and adapts them to global models. 


\section{Future Research Direction}
\label{section6}

Despite the intensive studies on FL, there still exists room for improvement. In this section, we suggest future research direction that needs more attention. \\

\noindent\textbf{Standardization} 
Multiple institutions across the globe use different languages, and each institution has its own convention to format data instances. It means that local models are trained on differently formatted data instances, thus might incurring unnecessary deviation. Hence, the local models should go through a proper preprocessing step to standardize different types and contents of data. \\[-5pt]

\noindent\textbf{Various Type of Data}
Data are often not limited to a single type; for example, EHR contains texts, images, or tables. Numerous papers are, however, limited in that they have tested their methods on simple image datasets, such as the variety of MNISTs or CIFAR-10. Thus, their superiority might not stand out on text or a combination of multi-types of datasets. Moreover, most papers on backdoor attacks had been only simulated on image classification tasks. More experiments with various types of datasets are needed for more practicality. \\[-5pt]

\noindent\textbf{Incentive Mechanism}
A sufficient incentive mechanism is needed as well. Large institutions are highly likely to have enough data (in terms of quality and quantity) to build a model that makes predictions with reasonable confidence. If the coordination deteriorates performance, security, or privacy, they have no reason to participate and contribute, consuming their resources or risking privacy. A robust data quality verification could be a possible solution for encouragement. By ensuring that variable data produces a more generalizable and performance-enhanced global model, they may be willing to join. \\[-5pt]

\noindent\textbf{Balancing Trade-off}
Although multiple efforts have tried to prevent sudden client dropout, privacy is not well-preserved at the moment. For example, a server inquires about clients' geolocation to form a homogeneous group for client selection to deal with system heterogeneity \cite{abdulrahman2020fedmccs}. Nishio et al. \cite{nishio2019client} limit their global model to a simple DNN structure at the cost of accuracy. More research should be conducted to find a harmonious balance between privacy and performance in applying FL to real-world practice. \\[-5pt]

\noindent\textbf{United and Non-orthogonal Defense techniques}
All defensive methods are not mutually exclusive; a single defense technique can thwart multiple attacks. Nevertheless, existing countermeasures are mostly studied separately and orthogonal \cite{zhou2021augmented}. For example, some encryption-based methods come at a high communication cost, causing communication bottlenecks, and perturbation-based methods sacrifice performance. Defense objectives should be congratulated for more practicality.

\section{Conclusion}
\label{section7}
Federated Learning has been suggested as a new paradigm for utilizing AI techniques in data-sensitive industries. To train a comparable AI model thus far, data have been collected centrally, putting privacy at risk and incurring substantial communication costs. Such data collection, however, is no longer possible due to stringent regulations. FL comes with the benefit of creating well-performing ML/DL models without data disclosure or collection as if being trained on extensive data. It did, however, enlarge attack surfaces and introduce new vulnerabilities. Immense research works have been conducted to mitigate the vulnerabilities, yet open challenges still exist prohibiting its practical application. We expect that with this survey on vulnerabilities, attacks, and defense, researchers would pay greater attention to the unmet needs.

\section*{Acknowledgement}
This research was supported by Healthcare AI Convergence Research \& Development Program through the National IT Industry Promotion Agency of Korea (NIPA) funded by Ministry of Science and ICT (No. S1601-20-1041).

\bibliographystyle{splncs04}
\bibliography{main.bbl}

\begin{thebibliography}{10}
\providecommand{\url}[1]{\texttt{#1}}
\providecommand{\urlprefix}{URL }
\providecommand{\doi}[1]{https://doi.org/#1}

\bibitem{abdulrahman2020fedmccs}
AbdulRahman, S., Tout, H., Mourad, A., Talhi, C.: Fedmccs: Multicriteria client
  selection model for optimal iot federated learning. IEEE Internet of Things
  Journal  \textbf{8}(6),  4723--4735 (2020)

\bibitem{awan2021contra}
Awan, S., Luo, B., Li, F.: Contra: Defending against poisoning attacks in
  federated learning. In: European Symposium on Research in Computer Security.
  pp. 455--475. Springer (2021)

\bibitem{bagdasaryan2020backdoor}
Bagdasaryan, E., Veit, A., Hua, Y., Estrin, D., Shmatikov, V.: How to backdoor
  federated learning. In: International Conference on Artificial Intelligence
  and Statistics. pp. 2938--2948. PMLR (2020)

\bibitem{baruch2019little}
Baruch, G., Baruch, M., Goldberg, Y.: A little is enough: Circumventing
  defenses for distributed learning. Advances in Neural Information Processing
  Systems  \textbf{32} (2019)

\bibitem{benmalek2022security}
Benmalek, M., Benrekia, M.A., Challal, Y.: Security of federated learning:
  Attacks, defensive mechanisms, and challenges. Revue des Sciences et
  Technologies de l'Information-S{\'e}rie RIA: Revue d'Intelligence
  Artificielle  \textbf{36}(1),  49--59 (2022)

\bibitem{bhagoji2019analyzing}
Bhagoji, A.N., Chakraborty, S., Mittal, P., Calo, S.: Analyzing federated
  learning through an adversarial lens. In: International Conference on Machine
  Learning. pp. 634--643. PMLR (2019)

\bibitem{bhushan2017man}
Bhushan, B., Sahoo, G., Rai, A.K.: Man-in-the-middle attack in wireless and
  computer networking—a review. In: 2017 3rd International Conference on
  Advances in Computing, Communication \& Automation (ICACCA)(Fall). pp.~1--6.
  IEEE (2017)

\bibitem{blanchard2017machine}
Blanchard, P., El~Mhamdi, E.M., Guerraoui, R., Stainer, J.: Machine learning
  with adversaries: Byzantine tolerant gradient descent. Advances in Neural
  Information Processing Systems  \textbf{30} (2017)

\bibitem{bouacida2021vulnerabilities}
Bouacida, N., Mohapatra, P.: Vulnerabilities in federated learning. IEEE Access
   \textbf{9},  63229--63249 (2021)

\bibitem{cao2020fltrust}
Cao, X., Fang, M., Liu, J., Gong, N.Z.: Fltrust: Byzantine-robust federated
  learning via trust bootstrapping. arXiv preprint arXiv:2012.13995  (2020)

\bibitem{chen2020zero}
Chen, Z., Tian, P., Liao, W., Yu, W.: Zero knowledge clustering based
  adversarial mitigation in heterogeneous federated learning. IEEE Transactions
  on Network Science and Engineering  \textbf{8}(2),  1070--1083 (2020)

\bibitem{fang2021privacy}
Fang, H., Qian, Q.: Privacy preserving machine learning with homomorphic
  encryption and federated learning. Future Internet  \textbf{13}(4), ~94
  (2021)

\bibitem{fang2020local}
Fang, M., Cao, X., Jia, J., Gong, N.: Local model poisoning attacks to
  $\{$Byzantine-Robust$\}$ federated learning. In: 29th USENIX Security
  Symposium (USENIX Security 20). pp. 1605--1622 (2020)

\bibitem{fraboni2021free}
Fraboni, Y., Vidal, R., Lorenzi, M.: Free-rider attacks on model aggregation in
  federated learning. In: International Conference on Artificial Intelligence
  and Statistics. pp. 1846--1854. PMLR (2021)

\bibitem{fredrikson2015model}
Fredrikson, M., Jha, S., Ristenpart, T.: Model inversion attacks that exploit
  confidence information and basic countermeasures. In: Proceedings of the 22nd
  ACM SIGSAC conference on computer and communications security. pp. 1322--1333
  (2015)

\bibitem{fung2020limitations}
Fung, C., Yoon, C.J., Beschastnikh, I.: The limitations of federated learning
  in sybil settings. In: 23rd International Symposium on Research in Attacks,
  Intrusions and Defenses (RAID 2020). pp. 301--316 (2020)

\bibitem{geiping2020inverting}
Geiping, J., Bauermeister, H., Dr{\"o}ge, H., Moeller, M.: Inverting
  gradients-how easy is it to break privacy in federated learning? Advances in
  Neural Information Processing Systems  \textbf{33},  16937--16947 (2020)

\bibitem{guerraoui2018hidden}
Guerraoui, R., Rouault, S., et~al.: The hidden vulnerability of distributed
  learning in byzantium. In: International Conference on Machine Learning. pp.
  3521--3530. PMLR (2018)

\bibitem{haddadpour2021federated}
Haddadpour, F., Kamani, M.M., Mokhtari, A., Mahdavi, M.: Federated learning
  with compression: Unified analysis and sharp guarantees. In: International
  Conference on Artificial Intelligence and Statistics. pp. 2350--2358. PMLR
  (2021)

\bibitem{hao2019efficient}
Hao, M., Li, H., Luo, X., Xu, G., Yang, H., Liu, S.: Efficient and
  privacy-enhanced federated learning for industrial artificial intelligence.
  IEEE Transactions on Industrial Informatics  \textbf{16}(10),  6532--6542
  (2019)

\bibitem{ho2021dp}
Ho, S., Qu, Y., Gu, B., Gao, L., Li, J., Xiang, Y.: Dp-gan: Differentially
  private consecutive data publishing using generative adversarial nets.
  Journal of Network and Computer Applications  \textbf{185},  103066 (2021)

\bibitem{huang2020fairness}
Huang, W., Li, T., Wang, D., Du, S., Zhang, J.: Fairness and accuracy in
  federated learning. arXiv preprint arXiv:2012.10069  (2020)

\bibitem{jeong2018communication}
Jeong, E., Oh, S., Kim, H., Park, J., Bennis, M., Kim, S.L.:
  Communication-efficient on-device machine learning: Federated distillation
  and augmentation under non-iid private data. arXiv preprint arXiv:1811.11479
  (2018)

\bibitem{jiang2022model}
Jiang, Y., Wang, S., Valls, V., Ko, B.J., Lee, W.H., Leung, K.K., Tassiulas,
  L.: Model pruning enables efficient federated learning on edge devices. IEEE
  Transactions on Neural Networks and Learning Systems  (2022)

\bibitem{hipaa_journal_2022}
Journal, H.: December 2021 healthcare data breach report (Jun 2022),
  \url{https://www.hipaajournal.com/december-2021-healthcare-data-breach-report/}

\bibitem{kanagavelu2020two}
Kanagavelu, R., Li, Z., Samsudin, J., Yang, Y., Yang, F., Goh, R.S.M., Cheah,
  M., Wiwatphonthana, P., Akkarajitsakul, K., Wang, S.: Two-phase multi-party
  computation enabled privacy-preserving federated learning. In: 2020 20th
  IEEE/ACM International Symposium on Cluster, Cloud and Internet Computing
  (CCGRID). pp. 410--419. IEEE (2020)

\bibitem{khosravy2022model}
Khosravy, M., Nakamura, K., Hirose, Y., Nitta, N., Babaguchi, N.: Model
  inversion attack by integration of deep generative models: Privacy-sensitive
  face generation from a face recognition system. IEEE Transactions on
  Information Forensics and Security  \textbf{17},  357--372 (2022).
  \doi{10.1109/TIFS.2022.3140687}

\bibitem{konevcny2016federated}
Kone{\v{c}}n{\`y}, J., McMahan, H.B., Yu, F.X., Richt{\'a}rik, P., Suresh,
  A.T., Bacon, D.: Federated learning: Strategies for improving communication
  efficiency. arXiv preprint arXiv:1610.05492  (2016)

\bibitem{li2019teev}
Li, W., Xia, Y., Lu, L., Chen, H., Zang, B.: Teev: Virtualizing trusted
  execution environments on mobile platforms. In: Proceedings of the 15th ACM
  SIGPLAN/SIGOPS International Conference on Virtual Execution Environments.
  pp. 2--16 (2019)

\bibitem{lin2019free}
Lin, J., Du, M., Liu, J.: Free-riders in federated learning: Attacks and
  defenses. arXiv preprint arXiv:1911.12560  (2019)

\bibitem{lu2020efficient}
Lu, Y., Fan, L.: An efficient and robust aggregation algorithm for learning
  federated cnn. In: Proceedings of the 2020 3rd International Conference on
  Signal Processing and Machine Learning. pp.~1--7 (2020)

\bibitem{lyu2020privacy}
Lyu, L., Yu, H., Ma, X., Sun, L., Zhao, J., Yang, Q., Yu, P.S.: Privacy and
  robustness in federated learning: Attacks and defenses. arXiv preprint
  arXiv:2012.06337  (2020)

\bibitem{lyu2020threats}
Lyu, L., Yu, H., Yang, Q.: Threats to federated learning: A survey. arXiv
  preprint arXiv:2003.02133  (2020)

\bibitem{ma2019data}
Ma, Y., Zhu, X., Hsu, J.: Data poisoning against differentially-private
  learners: Attacks and defenses. arXiv preprint arXiv:1903.09860  (2019)

\bibitem{mallik2019man}
Mallik, A.: Man-in-the-middle-attack: Understanding in simple words.
  Cyberspace: Jurnal Pendidikan Teknologi Informasi  \textbf{2}(2),  109--134
  (2019)

\bibitem{melis2019exploiting}
Melis, L., Song, C., De~Cristofaro, E., Shmatikov, V.: Exploiting unintended
  feature leakage in collaborative learning. In: 2019 IEEE symposium on
  security and privacy (SP). pp. 691--706. IEEE (2019)

\bibitem{miao2022against}
Miao, L., Yang, W., Hu, R., Li, L., Huang, L.: Against backdoor attacks in
  federated learning with differential privacy. In: ICASSP 2022-2022 IEEE
  International Conference on Acoustics, Speech and Signal Processing (ICASSP).
  pp. 2999--3003. IEEE (2022)

\bibitem{mo2019efficient}
Mo, F., Haddadi, H.: Efficient and private federated learning using tee. In:
  Proc. EuroSys Conf., Dresden, Germany (2019)

\bibitem{mo2021ppfl}
Mo, F., Haddadi, H., Katevas, K., Marin, E., Perino, D., Kourtellis, N.: Ppfl:
  privacy-preserving federated learning with trusted execution environments.
  In: Proceedings of the 19th Annual International Conference on Mobile
  Systems, Applications, and Services. pp. 94--108 (2021)

\bibitem{mothukuri2021survey}
Mothukuri, V., Parizi, R.M., Pouriyeh, S., Huang, Y., Dehghantanha, A.,
  Srivastava, G.: A survey on security and privacy of federated learning.
  Future Generation Computer Systems  \textbf{115},  619--640 (2021)

\bibitem{nasr2019comprehensive}
Nasr, M., Shokri, R., Houmansadr, A.: Comprehensive privacy analysis of deep
  learning: Passive and active white-box inference attacks against centralized
  and federated learning. In: 2019 IEEE symposium on security and privacy (SP).
  pp. 739--753. IEEE (2019)

\bibitem{nguyen2021flame}
Nguyen, T.D., Rieger, P., Chen, H., Yalame, H., M{\"o}llering, H., Fereidooni,
  H., Marchal, S., Miettinen, M., Mirhoseini, A., Zeitouni, S., et~al.: Flame:
  Taming backdoors in federated learning. Cryptology ePrint Archive  (2021)

\bibitem{nishio2019client}
Nishio, T., Yonetani, R.: Client selection for federated learning with
  heterogeneous resources in mobile edge. In: ICC 2019-2019 IEEE international
  conference on communications (ICC). pp.~1--7. IEEE (2019)

\bibitem{pan2020privacy}
Pan, X., Zhang, M., Ji, S., Yang, M.: Privacy risks of general-purpose language
  models. In: 2020 IEEE Symposium on Security and Privacy (SP). pp. 1314--1331.
  IEEE (2020)

\bibitem{parisot2021property}
Parisot, M.P.M., Pejo, B., Spagnuelo, D.: Property inference attacks on
  convolutional neural networks: Influence and implications of target model's
  complexity. CoRR  \textbf{abs/2104.13061} (2021),
  \url{https://arxiv.org/abs/2104.13061}

\bibitem{shejwalkar2021manipulating}
Shejwalkar, V., Houmansadr, A.: Manipulating the byzantine: Optimizing model
  poisoning attacks and defenses for federated learning. In: NDSS (2021)

\bibitem{shokri2017membership}
Shokri, R., Stronati, M., Song, C., Shmatikov, V.: Membership inference attacks
  against machine learning models. In: 2017 IEEE Symposium on Security and
  Privacy (SP). pp. 3--18 (2017). \doi{10.1109/SP.2017.41}

\bibitem{sun2019can}
Sun, Z., Kairouz, P., Suresh, A.T., McMahan, H.B.: Can you really backdoor
  federated learning? arXiv preprint arXiv:1911.07963  (2019)

\bibitem{truex2019hybrid}
Truex, S., Baracaldo, N., Anwar, A., Steinke, T., Ludwig, H., Zhang, R., Zhou,
  Y.: A hybrid approach to privacy-preserving federated learning. In:
  Proceedings of the 12th ACM workshop on artificial intelligence and security.
  pp. 1--11 (2019)

\bibitem{wang2020attack}
Wang, H., Sreenivasan, K., Rajput, S., Vishwakarma, H., Agarwal, S., Sohn,
  J.y., Lee, K., Papailiopoulos, D.: Attack of the tails: Yes, you really can
  backdoor federated learning. Advances in Neural Information Processing
  Systems  \textbf{33},  16070--16084 (2020)

\bibitem{wang2022flare}
Wang, N., Xiao, Y., Chen, Y., Hu, Y., Lou, W., Hou, Y.T.: Flare: Defending
  federated learning against model poisoning attacks via latent space
  representations. In: Proceedings of the 2022 ACM on Asia Conference on
  Computer and Communications Security. pp. 946--958 (2022)

\bibitem{wu2020mitigating}
Wu, C., Yang, X., Zhu, S., Mitra, P.: Mitigating backdoor attacks in federated
  learning. arXiv preprint arXiv:2011.01767  (2020)

\bibitem{wu2022communication}
Wu, C., Wu, F., Lyu, L., Huang, Y., Xie, X.: Communication-efficient federated
  learning via knowledge distillation. Nature communications  \textbf{13}(1),
  ~1--8 (2022)

\bibitem{xie2019dba}
Xie, C., Huang, K., Chen, P.Y., Li, B.: Dba: Distributed backdoor attacks
  against federated learning. In: International Conference on Learning
  Representations (2019)

\bibitem{yin2018byzantine}
Yin, D., Chen, Y., Kannan, R., Bartlett, P.: Byzantine-robust distributed
  learning: Towards optimal statistical rates. In: International Conference on
  Machine Learning. pp. 5650--5659. PMLR (2018)

\bibitem{zhang2020batchcrypt}
Zhang, C., Li, S., Xia, J., Wang, W., Yan, F., Liu, Y.: $\{$BatchCrypt$\}$:
  Efficient homomorphic encryption for $\{$Cross-Silo$\}$ federated learning.
  In: 2020 USENIX annual technical conference (USENIX ATC 20). pp. 493--506
  (2020)

\bibitem{zhang2022fine}
Zhang, L., Shen, L., Ding, L., Tao, D., Duan, L.Y.: Fine-tuning global model
  via data-free knowledge distillation for non-iid federated learning. In:
  Proceedings of the IEEE/CVF Conference on Computer Vision and Pattern
  Recognition. pp. 10174--10183 (2022)

\bibitem{zhang2020secret}
Zhang, Y., Jia, R., Pei, H., Wang, W., Li, B., Song, D.: The secret revealer:
  Generative model-inversion attacks against deep neural networks. In:
  Proceedings of the IEEE/CVF conference on computer vision and pattern
  recognition. pp. 253--261 (2020)

\bibitem{zhou2021augmented}
Zhou, Z., Xu, C., Wang, M., Ma, T., Yu, S.: Augmented dual-shuffle-based moving
  target defense to ensure cia-triad in federated learning. In: 2021 IEEE
  Global Communications Conference (GLOBECOM). pp. 01--06. IEEE (2021)

\end{thebibliography}

\end{document}